\documentclass[final,3p,times,twocolumn]{elsarticle}

\usepackage{amssymb}
\usepackage{amsmath}

\usepackage[normalem]{ulem}
\usepackage[hidelinks, colorlinks]{hyperref}

\usepackage{booktabs} 

\usepackage{lineno}

\journal{Computers and Graphics}

\begin{document}

\begin{frontmatter}

\title{Style-Aware Gloss Control for Generative Non-Photorealistic Rendering}

\author{Santiago Jimenez-Navarro\corref{cor1}}
\ead{s.jimenez@unizar.es}
\ead[url]{https://santiagojn.github.io/}

\author{Belen Masia}
\ead{bmasia@unizar.es}
\ead[url]{https://webdiis.unizar.es/~bmasia/}

\author{Ana Serrano}
\ead{anase@unizar.es}
\ead[url]{https://ana-serrano.github.io/}

\cortext[cor1]{Corresponding Author. Address: University of Zaragoza, Department of Computer Science and Systems Engineering (DIIS), Zaragoza, Spain}

\affiliation{organization={University of Zaragoza -- I3A},
            country={Spain}}

\begin{abstract} %
Humans can infer material characteristics of objects from their visual appearance, and this ability extends to artistic depictions, where similar perceptual strategies guide the interpretation of paintings or drawings.
Among the factors that define material appearance, gloss, along with color, is widely regarded as one of the most important, and recent studies indicate that humans can perceive gloss independently of the artistic style used to depict an object.
To investigate how gloss and artistic style are represented in learned models, we train an unsupervised generative model on a newly curated dataset of painterly objects designed to systematically vary such factors.
Our analysis reveals a hierarchical latent space in which gloss is disentangled from other appearance factors, allowing for a detailed study of how gloss is represented and varies across artistic styles.
Building on this representation, we introduce a lightweight adapter that connects our style- and gloss-aware latent space to a latent-diffusion model, enabling the synthesis of non-photorealistic images with fine-grained control of these factors.
We compare our approach with previous models and observe improved disentanglement and controllability of the learned factors.
\end{abstract}


\end{frontmatter}

\begin{figure*}[t]
    \centering
    \includegraphics[width=0.95\linewidth]{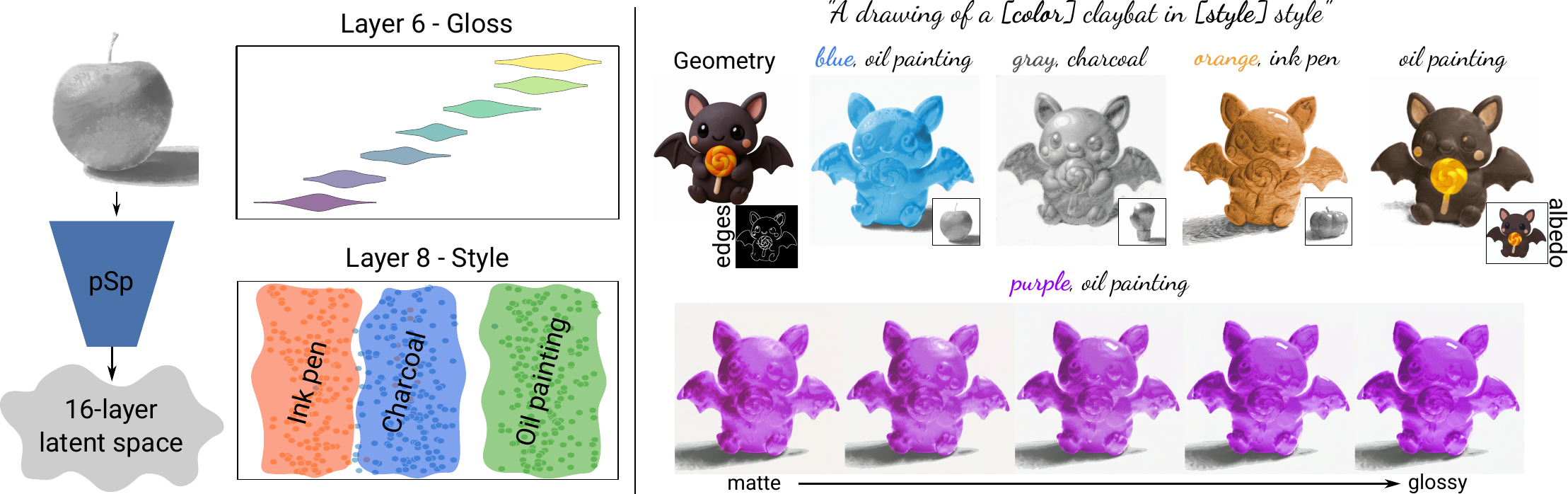}
    \caption{
        \emph{Left:} We train a pSp–StyleGAN2 pipeline that learns a 16-layer latent space where appearance factors such as gloss and artistic style emerge in a disentangled and hierarchical manner. Specific layers specialize in different attributes, e.g., Layer 6 captures gloss and Layer 8 captures style.
        \emph{Right:} This learned space enables intuitive control of appearance.
        Given a reference geometry (edges) and optional albedo map, the user can employ our diffusion-based pipeline to transfer the gloss and style of an input drawing (inset) to new objects, or traverse the gloss dimension to obtain predictable variations from matte to glossy while keeping other factors stable.
    }
    \label{fig:teaser}
\end{figure*}

\section{Introduction}

Visual perception plays a fundamental role in everyday life. We constantly evaluate the appearance of surrounding objects in order to interact with them effectively, either by not stepping on a wet floor, or by not touching a sharp blade. While such processing is largely automatic for humans, the perceived appearance of materials arises from a complex interplay of factors such as shape, illumination, and motion~\cite{fleming2003real, fleming2017material, schmidt2025core}. This interaction is not yet fully understood, and remains an active area of research~\cite{lagunas2021joint, serrano2021effect, chen2021effect}. For example, the same golden ring may appear shinier under the bright spotlights of a jewelry display than under dull lighting on a tabletop. 
Beyond photorealistic scenes, this 
is also relevant in 
non-photorealistic renditions %
(e.g., paintings), where an additional factor may influence appearance perception: 
the style with which they are depicted.
Among the different dimensions of material appearance, gloss is one of the most relevant attributes, together with color, and has therefore received particular attention in perception studies~\cite{chadwick2015perception, fleming2017material}. Recent studies suggest that the human brain is able to disambiguate gloss from style, relying on similar visual cues across different media~\cite{zuijlen2020painterly, delanoy2021perception, zhao2024material}. Because gloss is a key component of material appearance and may be influenced by the painting style, it is a particularly relevant attribute for studying appearance in non-photorealistic images.
In this work, we investigate whether a generative network trained on stylized renditions of objects can, without supervision, learn a hierarchical latent representation in which gloss is separated from other appearance factors such as style (\autoref{sec:space}).
This representation allows direct analysis of how visual factors are internally organized (\autoref{sec:analysis}), and provides a basis for controlled manipulation of gloss in images.

Building on the learnt representation, we next explore its potential for image synthesis (\autoref{sec:diffusion}). Generative models based on latent diffusion denoising processes are, together with flow-based models, the current state of the art in image synthesis~\cite{po2024state}.
While these models can produce high-quality images, 
a fine-grained control of the output
is difficult, and represents an active area of research~\cite{zhang2023adding,bhat2024loosecontrol,li2024controlnet++,bernal2025precisecam}. 
Conversely, GAN-based models have been extensively studied in the literature, where their favorable latent-space properties (e.g., interpretability, traversability, and continuity) have been exploited to identify editing directions that enable an intuitive manipulation of high-level image features~\cite{harkonen2020ganspace,bau2019gandissect}. However, these approaches are inherently limited by the reduced generative capacity of the GAN's generator.
Inspired by prior work on controlled generation~\cite{li2024stylegan, gandikota2024concept,jimenez2025controllable,gandikota2025sliderspace}, we explore how our latent space can be exploited to enable precise manipulation of style and gloss within diffusion models, two appearance factors that are normally hard to define through text prompts alone (\autoref{sec:diffusion}, \autoref{fig:traversalsSOTA}).

Overall, our latent representation contains an interpretable and disentangled encoding of gloss and other perceptual factors present in stylized images and enables fine-grained control of a diffusion pipeline.
Thus, our main contribution is a detailed analysis of how style influences the internal representation of visual attributes within a hierarchical, learning-based network trained without explicit supervision, revealing the emergence of a dedicated gloss dimension. Building on this analysis, we leverage the learned representation to enable diffusion-based image synthesis with fine-grained appearance control, achieving a level of gloss manipulation that is not attainable with existing style-transfer or general-purpose diffusion methods. 

Our trained model and code, together with our dataset, comprised of non-photorealistic renditions of objects depicted under different styles and levels of gloss, can be accessed at \url{https://graphics.unizar.es/projects/JimenezNavarro2026-NPRdisentanglement/}.

\section{Related Work}

\subsection{Image Stylization in Non-photorealistic Rendering}

Non-photorealistic rendering (NPR) refers to computer-graphics techniques that deliberately depart from photorealism to produce images with a hand-crafted or illustrative appearance.  
Rather than reproducing the physical accuracy of light transport, NPR methods aim to convey the content of a scene through artistic or stylized depiction.  
A central task in this area is image stylization, where a source image is transformed to match a chosen artistic style.
Early approaches address this task by explicitly defining brushstrokes or procedural painting rules.  
While effective, these analytical methods are typically slow, non-intuitive, and require manual parameter tuning~\cite{haeberli1990paint,hertzmann1998painterly}, which limits usability~\cite{hertzmann2003survey}.  

Since the introduction of Neural Style Transfer by Gatys et al.~\cite{gatys2015neural}, a variety of approaches have been developed to automatically apply a reference style to an image. Early CNN-based methods~\cite{gatys2015neural,jing2018stroke} rely on high-level features extracted from pretrained networks such as VGG~\cite{simonyan2014veryvgg}. Later, StyleGAN-based models employed the classic \textit{min–max} training of generator and discriminator to produce more realistic and diverse results~\cite{abdal2019image2stylegan,tang2021attentiongan}. 
Most recently, diffusion-based models have achieved state-of-the-art performance in terms of image quality, robustness, and diversity~\cite{qi2024deadiff,wang2025styleadapter,lei2025stylestudio,jiang2025balanced}, including training-free self-attention KV manipulation~\cite{Chung_2024_CVPR} and the identification of transformer blocks that predominantly encode style features~\cite{wang2024instantstyle}. 
A key challenge of this last type of models is their limited controllability, which has been mitigated with reasonable success by auxiliary networks that allow finer conditioning during synthesis~\cite{zhang2023adding,bhat2024loosecontrol,li2024controlnet++,bernal2025precisecam,wang2025lineart}.
For a more detailed discussion on the taxonomy and evolution of the aforementioned methods, we refer the reader to the survey by Wang et al.~\cite{wang2024learning}.

In our work, we condition a diffusion model on artistic styles represented in the $W+$ latent space of StyleGAN, a representation that offers both expressivity and controllability~\cite{li2024stylegan,gandikota2024concept}.  
Our focus is on capturing style and gloss levels independently and simultaneously, extending prior efforts in NPR stylization.  
The closest related work is Subias et al.~\cite{subias2025artistinator}, who introduced a diffusion pipeline for text-driven stylization. Their approach improves stylization flexibility by leveraging the generative capabilities of the vision-language model CLIP~\cite{radford2021learning}. However, it does not explicitly disentangle appearance factors and therefore cannot provide the predictable, continuous control enabled by our method.
In contrast, our approach uses unsupervised training and a disentangled $W+$ latent space to represent generative factors. This yields greater interpretability than the CLIP-based conditioning used by Subias et al., enabling continuous control over gloss rather than relying on discrete gloss categories, while also requiring only a fraction of the training data (10,080 vs. 1,336,272 samples).

\subsection{Perception in Non-photorealistic Rendering}
\label{sec:RW:NPR_perception}

Artistic depictions have long served as a tool for studying human visual perception. To represent real scenes, artists frequently use deliberate abstractions that allow them to effectively convey visual attributes of objects without adhering to strict photorealism~\cite{delanoy2021perception, zuijlen2020painterly, cavanagh2005artist}. 
Although some early studies reported differing interpretations~\cite{bousseau2013gloss}, more recent work supports the view that the human visual system relies on similar cues to judge material properties across different 
media~\cite{zuijlen2020painterly, delanoy2021perception, zhao2024material}, allowing systematic studies to extend beyond strictly photorealistic images.

Non-photorealistic depictions have been used to investigate a variety of material properties, including transparency~\cite{sayim2011art}, translucency~\cite{wijntjes2020thurstonian}, and specific materials such as fabrics~\cite{di2021soft,thomas1994fabric}.
Among these attributes, gloss has received particular attention 
in non-photorealistic settings
~\cite{delanoy2021perception,cavanagh2008reflections,di2019understanding,subias2025artistinator}, since gloss, together with color, is widely regarded as one of the most important attributes of material appearance~\cite{chadwick2015perception,fleming2017material}. 
In this work, we investigate how attributes related to different painting techniques--charcoal, oil painting, and ink pen--are naturally organized in the inner layers of a generative model.
The closest studies on this aspect are those of Zhao et al.~\cite{zhao2023zooming} and Elgammal et al.~\cite{elgammal2018shape}. However, both of them focus on interpreting a low-dimensional embedding space and its relation to specific painters or creation years, rather than studying how information about painting techniques can be captured within a hierarchical generator.

\subsection{Generative Models and Human Perception}
Humans excel at estimating physical properties of materials by sight. This core ability allows us to interact effectively with our surroundings, assessing whether a certain object is edible, slippery, or sharp~\cite{fleming2013perceptual, fleming2017material, schmidt2025core}. This material perception results from a complex interplay between intrinsic and extrinsic factors~\cite{fleming2003real, lagunas2021joint}. However, it is argued that the brain does not explicitly recover the physical reflectance parameters of a surface, but instead relies on statistical generative representations to perceive material appearance~\cite{fleming2014visual}. 

Learning-based generative models are increasingly being used to investigate the neural mechanisms underlying human visual perception~\cite{zhuang2021ventral,raugel2025disentangling}.
As discussed in \autoref{sec:RW:NPR_perception}, gloss perception has been a common benchmark for assessing the correspondence between artificial networks and human visual judgments~\cite{storrs2021unsupervised,prokott2021gloss,morimoto2025human}. Other factors, such as translucency~\cite{liao2023unsupervised}, viscosity~\cite{van2020visual} and refractive or reflective properties~\cite{tamura2022distinguishing} have also been explored in this context.
Demonstrating that these models can simulate human perception requires more than reproducing correct judgments; it also entails showing that they exhibit similar failure patterns.
Building on this idea, a growing line of work uses generative models to study visual illusions, examining how the visual system can be misled by specific image statistics~\cite{watanabe2018illusory,jaini2024intriguing,gomez2025art}.
Such studies help identify the internal mechanisms that lead to systematic misperceptions and can even guide the creation of new illusions~\cite{gomez2025art}.

In this work, we examine the features that naturally emerge within a StyleGAN2-ADA generator~\cite{karras2020trainingsgan2ada} and analyze how their internal organization relates to 
visual representations of perceptual features.

\section{Disentanglement of Style and Gloss in W+ space}
\label{sec:space}

Our goal is to investigate how gloss is represented in a generative model specialized for non-photorealistic rendering across a set of styles. Building on the incremental process of hierarchical generators such as StyleGAN2~\cite{karras2020analyzingsgan2}, we evaluate whether interpretable and informative dimensions of material appearance can naturally emerge in a model trained without explicit supervision and, in that case, how and why they are organized in that manner. 
While the use of multiple artistic styles introduces arbitrariness in the representations (e.g., of a given gloss level), it also allows us to study if factors that evoke the sense of gloss can be shared across different styles.

\subsection{Preliminaries}
\label{sec:preliminaries}

We rely on an unsupervised learning-based approach to study how gloss perception is affected by artistic styles in non-photorealistic rendering. In this section, we give background on the neural architectures that form the final pipeline, and the dataset specially created for this task.

Unsupervised approaches have proven to be useful for providing insights on how the human visual system may perceive complex stimuli from the real world~\cite{FLEMING2019100, storrs2021unsupervised}. In this work, we use a version of the well known Generative Adversarial Network (GAN)~\cite{goodfellow2014generative}, StyleGAN2-ADA~\cite{karras2020trainingsgan2ada}, together with a style encoder~\cite{richardson2021encodingpsp} to easily project samples into the latent space.

\noindent\textbf{StyleGAN2-ADA.}
The StyleGAN architecture~\cite{karras2019stylesgan} represented a major step forward on the interpretability and explainability of GAN-synthesized images by redesigning the generator component. Authors propose embedding the traditional, normally-distributed latent representation $z \in Z$ to an intermediate $w \in W$, which is more suitable to represent the distribution of input data, allowing a more disentangled representation of the factors of variation. This is achieved by a learned non-linear mapping network $f: Z \rightarrow W$, where the resulting $w$ is injected to each convolutional layer of the synthesis network via affine transformations.

StyleGAN2~\cite{karras2020analyzingsgan2} revised the synthesis architecture of its predecessor to remove some characteristic artifacts, improve general image quality, and facilitate the inversion task~\cite{xia2022gan}. Finally, StyleGAN2-ADA~\cite{karras2020trainingsgan2ada} introduces an Adaptive Discriminator Augmentation mechanism that allows training a model from scratch with only \textit{a few thousand training images}.

In this work we use the StyleGAN2-ADA architecture~\cite{karras2020trainingsgan2ada} (henceforth denoted as StyleGAN2 for simplicity) %
for the aforementioned benefits in terms of disentanglement and interpretability of results. Furthermore, ADA improves training stability when data are scarce, reducing the need to gather a very large set of high-quality painterly depictions. %

\noindent\textbf{pixel2style2pixel (pSp) encoder.}
GAN-based methods have shown a great capability to generate high-quality and diverse images in an unconditioned way~\cite{xia2022gan}, closely approximating the distribution of datasets of varied size and complexity~\cite{kang2023scaling}. 
Furthermore, it has been shown that information tends to naturally organize in a semantically-meaningful way throughout the different hierarchical levels of the generator network~\cite{Jahanian2020On, shen2020interpreting}. This opens the door to an intuitive and predictable editing of some interpretable factors of the GAN-generated images, by traversing the latent space $Z$ (or any of its variants)~\cite{xingang2023drag, wu2021stylespace}.

In order to perform this editing with an unknown image $x$, it is first necessary to estimate how $x$ is represented in the latent space, which can be formulated as the function $I: x \rightarrow z^*$. This constitutes a complex inverse problem due to the non-convexity of the generator network. The literature contains an extensive number of approaches to this GAN inversion problem, which can be categorized as (i) optimization-based, where $z^*$ is estimated through a costly but precise optimization process~\cite{abdal2019image2stylegan},
(ii) learning-based techniques which create an auxiliary encoder network that approximates $I$~\cite{richardson2021encodingpsp, Omer2021e4e}, similar to an autoencoder~\cite{bank2023autoencoders}, and (iii) hybrid methods that combine both~\cite{zhu2016generative}.
We use a learning-based approach, namely pSp~\cite{richardson2021encodingpsp}, since it is able to perform inversion in a single forward pass, and it has been shown that these models can achieve better disentanglement and informativeness capabilities than the original $Z$~\cite{wu2021stylespace}. The pSp encoder expands the original StyleGAN's $W$ space (single 512-dimensional embeddings) to an extended $W+$ space (one independent 512-dimensional style vector per generator layer), using a small fully convolutional network, \textit{map2style}, which gradually projects a [D, D, 512] feature map, where D denotes the spatial resolution of the feature map at a given stage of the encoder, into a [1, 1, 512] $w_i$ vector~\cite{richardson2021encodingpsp}. A schematic illustration of the aforementioned $Z$, $W$, and $W+$ latent spaces can be visualized in the supplementary material (Sec. S2.1).

\subsection{A Dataset for Stylized Gloss}
\label{sec:dataset}

\begin{figure}[t]
    \centering
    \includegraphics[width=\columnwidth]{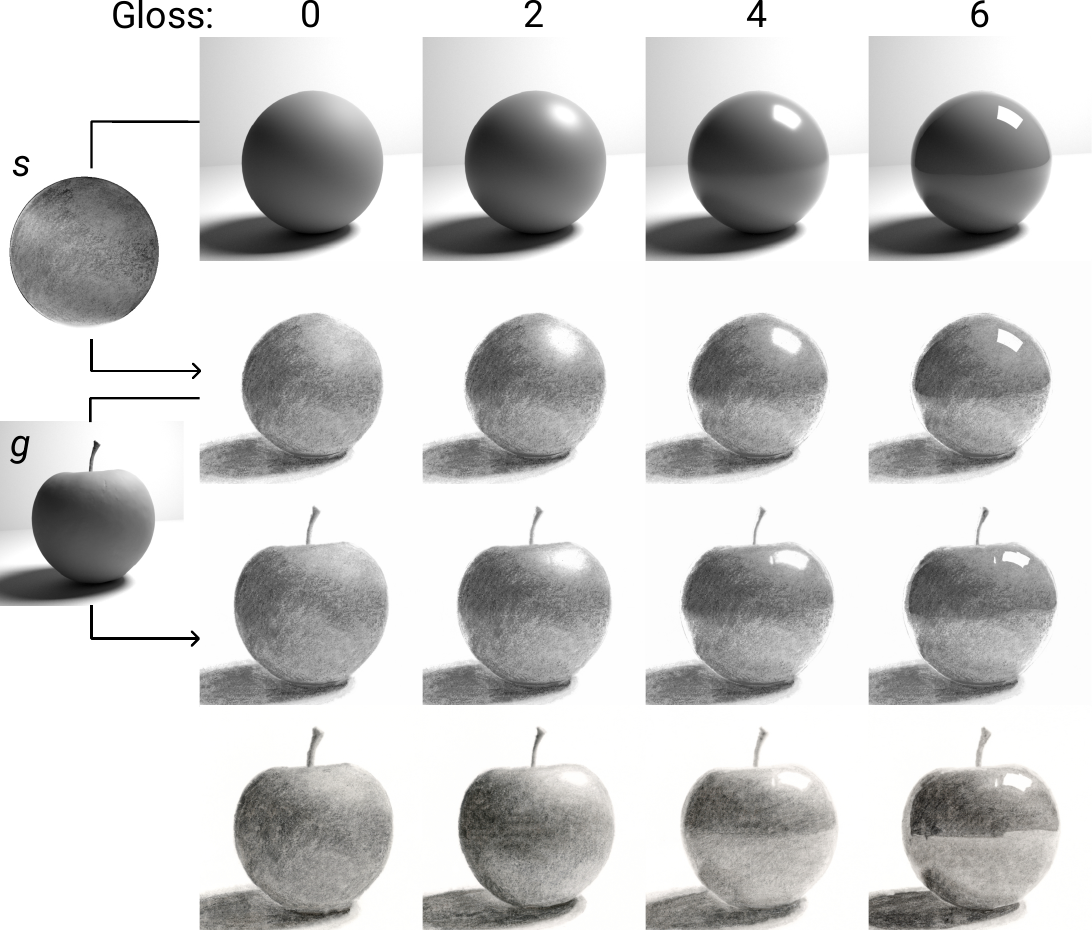}
    \caption{
        Application of the brushstroke map \textbf{s} to generate controlled style-guidance samples.     
        From top to bottom rows, we show 1) the photorealistic reference spheres with gradually varying roughness; 2) the stylized spheres using the estimated brushstroke map \textbf{s}; 3) the result of applying StyLit to the reference geometry \textbf{g}; and 4) the corresponding samples from the original Subias et al. dataset~\cite{subias2025artistinator}. Note the more controlled and continuous increment of gloss of our samples when compared with the reference. %
    }
    \label{fig:3:stylization}
\end{figure}

Although datasets for studying glossiness are common in photorealistic scenarios, their availability in NPR is limited. Existing datasets either offer limited control over generative factors~\cite{di2019understanding,van2021materials} or are too small to train an unsupervised generative model~\cite{delanoy2021perception}.
A recent work from Subias et al.~\cite{subias2025artistinator} introduced a varied and relatively controlled dataset of 1,336,272 painterly depictions generated with StyLit~\cite{fivser2016stylit}. 
The creation of this dataset involved two stages.
First, Subias et al. produced an original collection of reference pairs consisting of physically rendered spheres with controlled gloss levels and their corresponding artist-painted versions in different artistic styles.
Second, they used these painted spheres as style exemplars to transfer the appearance to a wide variety of geometries with StyLit, yielding the final large-scale set.
While the dataset is large-scale and offers controlled variability in generative factors, its direct use represents a major issue in our setup: because different gloss levels in the references come from different hand-painted spheres, a trained unsupervised model is likely to learn differences in brushstrokes rather than actual gloss variations (\autoref{fig:3:stylization}). Indeed, a preliminary experiment with our architecture showed early signs of clustering images by stroke patterns rather than gloss level. Therefore, for our purposes, appearance references (painted spheres) in a given style must be invariant with respect to brushstroke patterns.

\begin{figure}[t]
    \centering
    \includegraphics[width=\columnwidth]{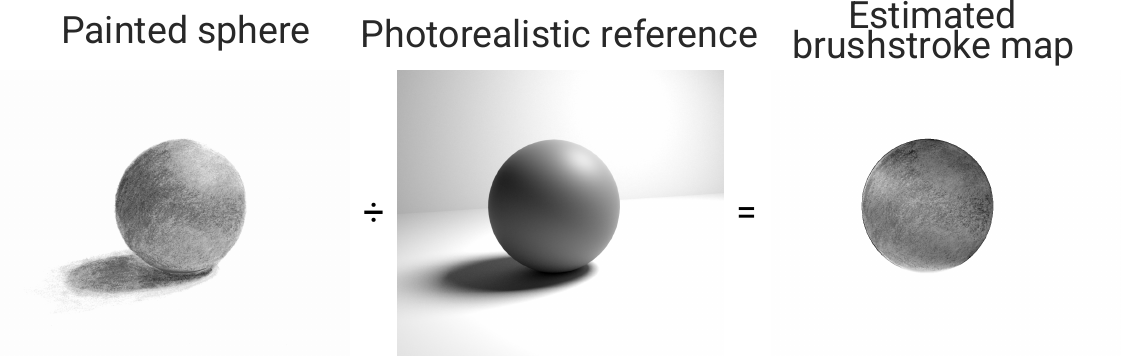}
    \caption{
        Example of brushstroke map extraction for a matte sphere painted in charcoal style.
        Artist-painted reference (\textit{left}), corresponding photorealistic rendering (\textit{middle}), and the estimated brushstroke map extracted from the pair (\textit{right}).    
    }
    \label{fig:3:brushstroke}
\end{figure}

For this purpose, we reprocess the reference pairs provided by Subias et al. (the physically rendered spheres with controlled gloss levels and their artist-painted counterparts). Inspired by the work of Litwinowicz~\cite{litwinowicz1997processing}, our goal is to create a brushstroke map \textbf{\textit{s}} that contains information about the artistic style. 
For each style, we select the painted sphere with the lowest gloss level, chosen to minimize specular highlights, because such highlights carry little information about the painting style. We then align it with its corresponding rendered sphere.
We compute \textbf{\textit{s}} by dividing the painted image by the rendered one in pixel space, applying constants and scaling factors to ensure numerical stability, and applying a mask to keep the foreground, discarding background and shadows (\autoref{fig:3:brushstroke}).
This isolated style representation is then applied to the rendered photorealistic spheres with different gloss levels (rendered with variations of the roughness \textit{r} parameter of the \textit{Disney's Principled BSDF}~\cite{burley2012physically}) and used as the style reference for StyLit~\cite{fivser2016stylit}. 
Each rendered sphere is produced with a known roughness value, providing an explicit gloss‐level label for every sample in the dataset. %
This yields stylized results of a given geometry \textbf{\textit{g}} with more controlled strokes than the original samples (\autoref{fig:3:stylization}). 
The final dataset results as the combination of three styles (charcoal, ink pen, and oil painting, which we believe are representative of painting styles), 20 geometries of varied complexity, four illuminations, seven gloss levels, and six colors, resulting in a dataset of 10,080 samples. These generative factors will be used later for evaluating the performance of the model trained without supervision. 
Representative samples of such dataset can be found in the supplementary material (Sec.~S1). 
To enable reproducibility, we share the code used in this dataset generation pipeline at \url{https://graphics.unizar.es/projects/JimenezNavarro2026-NPRdisentanglement/}.

\subsection{Architecture}
\label{sec:architecture}

\begin{figure}[t]
    \centering
    \includegraphics[width=\columnwidth]{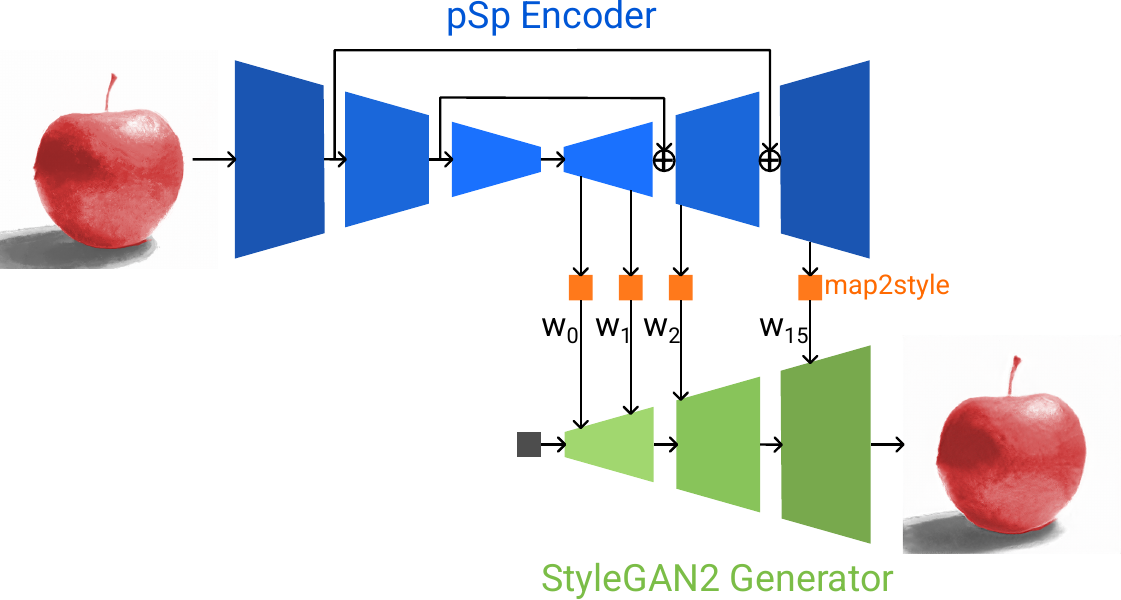}
    \caption{
        Diagram of the architecture composed by the pSp encoder that constructs a layer-wise latent space through \textit{map2style} layers, and the StyleGAN2 generator, which synthesizes images conditioned on these layer-wise latent representations.
    }
    \label{fig:diagram}
\end{figure}

The architecture (\autoref{fig:diagram}) is divided into two modules, trained separately: a StyleGAN2 generator, and a pSp encoder (\autoref{sec:preliminaries}). The generator learns to produce images resembling those of the training dataset (\autoref{sec:dataset}), through the traditional adversarial \textit{min–max} optimization of a generator and discriminator, where the latent space $W$ emerges. It also incorporates \textit{path length regularization}, introduced in StyleGAN2, to encourage smoothness in the latent space. Keeping the resulting generator frozen, a pSp encoder is trained for image reconstruction, creating a layer-wise latent space $W+$ where images can be projected in a single forward pass.
Its loss function $ \mathcal{L}(x) = \lambda_1 \mathcal{L}_2(x) + \lambda_2 \mathcal{L}_{LPIPS}(x) +\lambda_3 \mathcal{L}_{reg}(x) $, as proposed by Richardson et al.~\cite{richardson2021encodingpsp}, encourages both disentanglement (third term) and reconstruction accuracy (first and second terms) in the extended latent space, properties that are essential to our task. 
By training both modules in an unsupervised fashion, we do not impose any predefined factors or ordering to be learned in the latent space; instead, these representations are organically learned during the optimization process, using only images as input.
For further implementation details, please refer to the supplementary material (Sec.~S2.1). 

\section{Analysis of the Model}
\label{sec:analysis}

This section presents a comprehensive evaluation of the trained generative pipeline (\autoref{sec:architecture}). We begin by assessing its reconstruction ability, by projecting an input sample into $W+$ and feeding the resulting embedding to the generator, in order to evaluate potential information loss within the pipeline (\autoref{sec:reconstruct}). Next, we analyze key properties of the learned latent space, including continuity, traversability, disentanglement, and informativeness (\autoref{sec:latent_space}, \autoref{sec:embed_categorical}, and \autoref{sec:embed_gloss}). Finally, we run an ablation study on the number of styles handled by the network (\autoref{sec:ablationstyles}), studying the robustness of the method when scaling the complexity of the dataset.
Additional analysis regarding the internal organization of the latent space and further visualizations can be found in the supplementary material (Sec. S4.2).

\begin{figure}[t]
    \centering
    \includegraphics[width=\columnwidth]{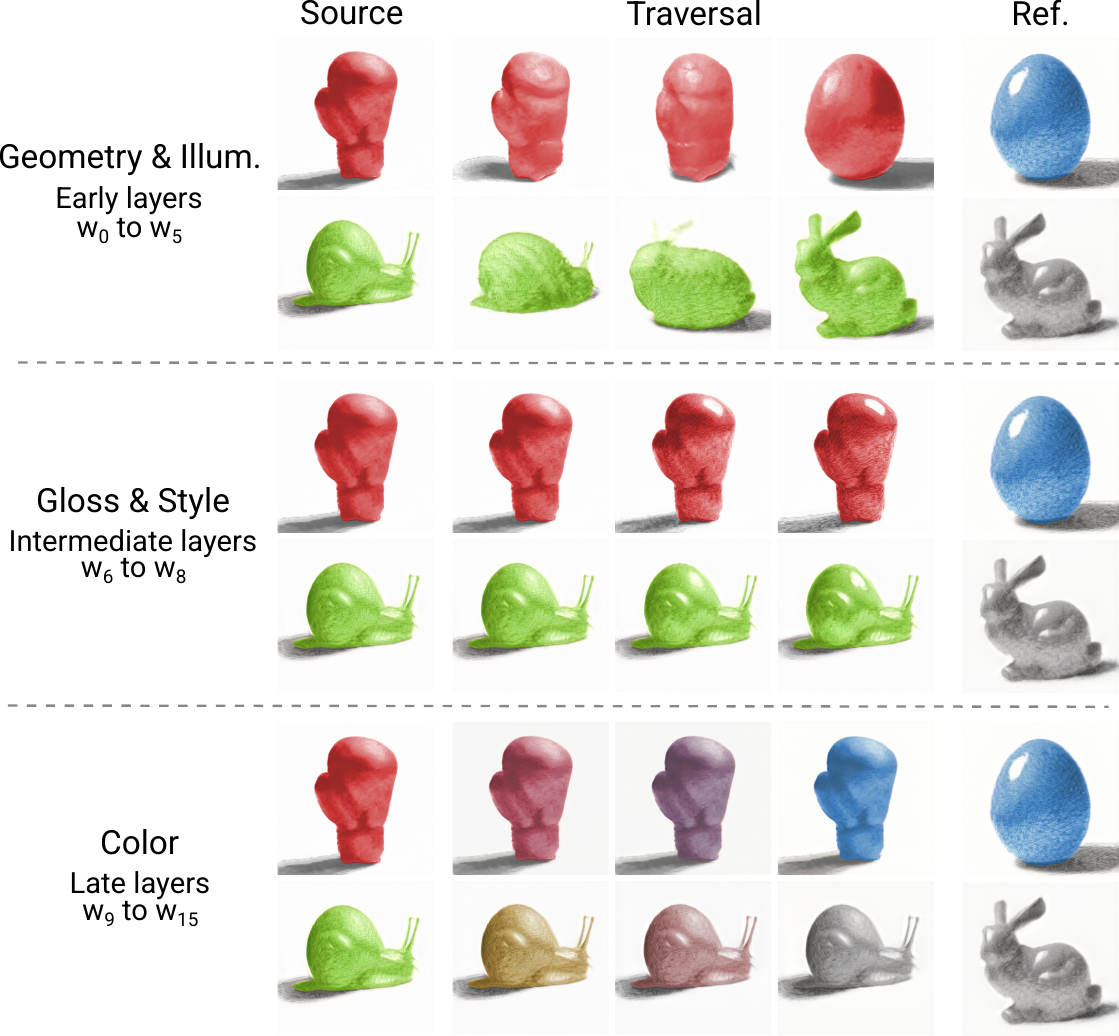}
    \caption{
        Traversals at different depths of the latent space produce interpretable changes in appearance.  
Starting from the embeddings of a source image (\textit{left}), moving along early (\textit{top}), intermediate (\textit{middle}), and late (\textit{bottom}) layers induces variations in geometry and illumination, gloss and style, or color, respectively.
    }
    \label{fig:traversals}
\end{figure}

\begin{figure*}[t]
    \centering
    \includegraphics[width=\textwidth]{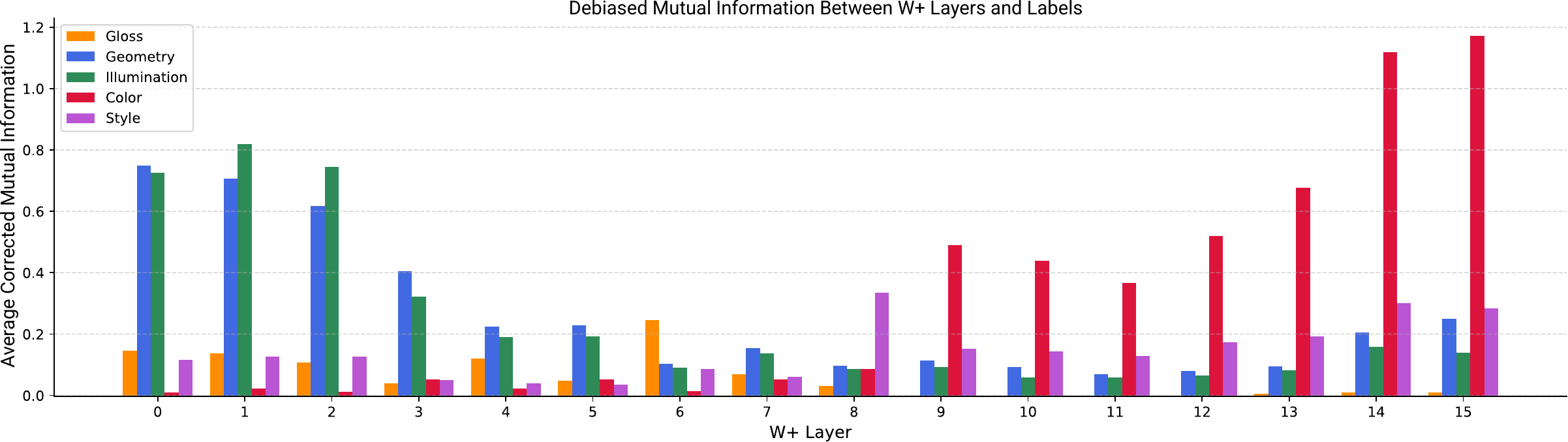}
    \caption{
        Per-layer corrected conditional mutual information scores between generative factors and $W+$ layers, illustrating how different attributes are encoded across the latent hierarchy after accounting for the influence of the remaining factors.
    }
    \label{fig:mi}
\end{figure*}

\subsection{Reconstruction Capabilities of the Pipeline}
\label{sec:reconstruct}
High reconstruction quality is an important prerequisite for obtaining a meaningful latent space.
Without it, any apparent disentanglement might reflect 
learned noise 
rather than true factor separation. To verify this, we evaluate whether generative features can propagate from the input image, encoded by pSp, through the StyleGAN2 generator. 
Our pipeline is able to
(i) accurately reconstruct general samples and (ii) closely reproduce the ground-truth gloss traversal (visual results of various reconstructions are available in the supplementary material, Sec.~S3). We also quantitatively evaluate the reconstruction capabilities of the pipeline by computing the metrics of MSE, MAE, PSNR, and SSIM between the reference and reconstructions, obtaining values of 0.003, 0.027, 25.51, and 0.801 respectively. %

\subsection{Internal Organization of the Latent Space}
\label{sec:latent_space}

When analyzing the structure of $W+$, we observed that information is hierarchically distributed across different layers of the space. We hypothesize that this organization arises from the specific information required by the StyleGAN generator at each stage of the synthesis process. As observed in \autoref{fig:traversals}, \textit{early layers} ($w_{0..5}$) primarily encode global scene attributes, such as geometry and illumination. The \textit{middle layers} ($w_{6..8}$) capture finer properties of material appearance, including gloss level and painting style. Finally, the \textit{late layers} ($w_{9..15}$) determine surface-level attributes such as color. Notably, we can observe a clear disentanglement among these factors, which enables predictable manipulations when traversing the latent space.

To quantify how different latent factors are encoded in the network, we exploit the explicit labels available from our dataset generation (gloss level, geometry, style, color, and illumination) and examine how this information is distributed across the 16 layers of the StyleGAN $W+$ latent space (each being 512-dimensional). 
For each layer, we compute a de-biased conditional mutual information~\cite{panzeri1996analytical,runge2019causal} between the layer’s embedding 
$X$ and a target factor 
$Y$ (e.g., gloss), while conditioning on the remaining factors 
$Z$ (e.g., geometry or style). 
This metric therefore quantifies how much information a StyleGAN layer encodes about a specific factor after accounting for the influence of all remaining factors and correcting for estimation bias.
Specifically, this measure subtracts the expected mutual information obtained when the labels $Y$ are randomly permuted, thereby correcting for spurious dependencies caused by correlations among factors:

\begin{align*}
Corr\_{MI}(X;Y|Z) &= I(X;Y|Z) - \mathbb{E}_{perm}[I(X;Y_{perm}|Z)], \\
\text{where } I(X;Y|Z) &= H(Y|Z) - H(Y|X,Z).
\end{align*}

Here, $I(X;Y|Z)$ is the conditional mutual information, and $H$ denotes entropy. $\mathbb{E}_{perm}$ represents the average mutual information arising from noise (estimated by permuting the labels), and subtracting it from the estimated conditional mutual information $I$ provides the corrected value $Corr\_MI$. 
\autoref{fig:mi} shows a complete visualization of the information distribution:
results of this quantitative analysis are coherent with the qualitative ones in \autoref{fig:traversals}, revealing that \emph{geometry} and \emph{illumination} are captured in layers 0 to 5, \emph{gloss} is captured in layer 6, \emph{style} is captured in layer 8, and \emph{color} is captured in layers 9 to 15.  
Additionally, the supplementary material contains an analysis on the image formation process by inspecting the intermediate representations used by the generator (Sec.~S4.2).

These results provide a clear, layer-wise mapping of how different appearance factors are organized inside the generator, offering an interpretable link between the latent architecture and the resulting visual attributes.

\begin{figure}[t]
    \centering
    \includegraphics[width=0.98\columnwidth]{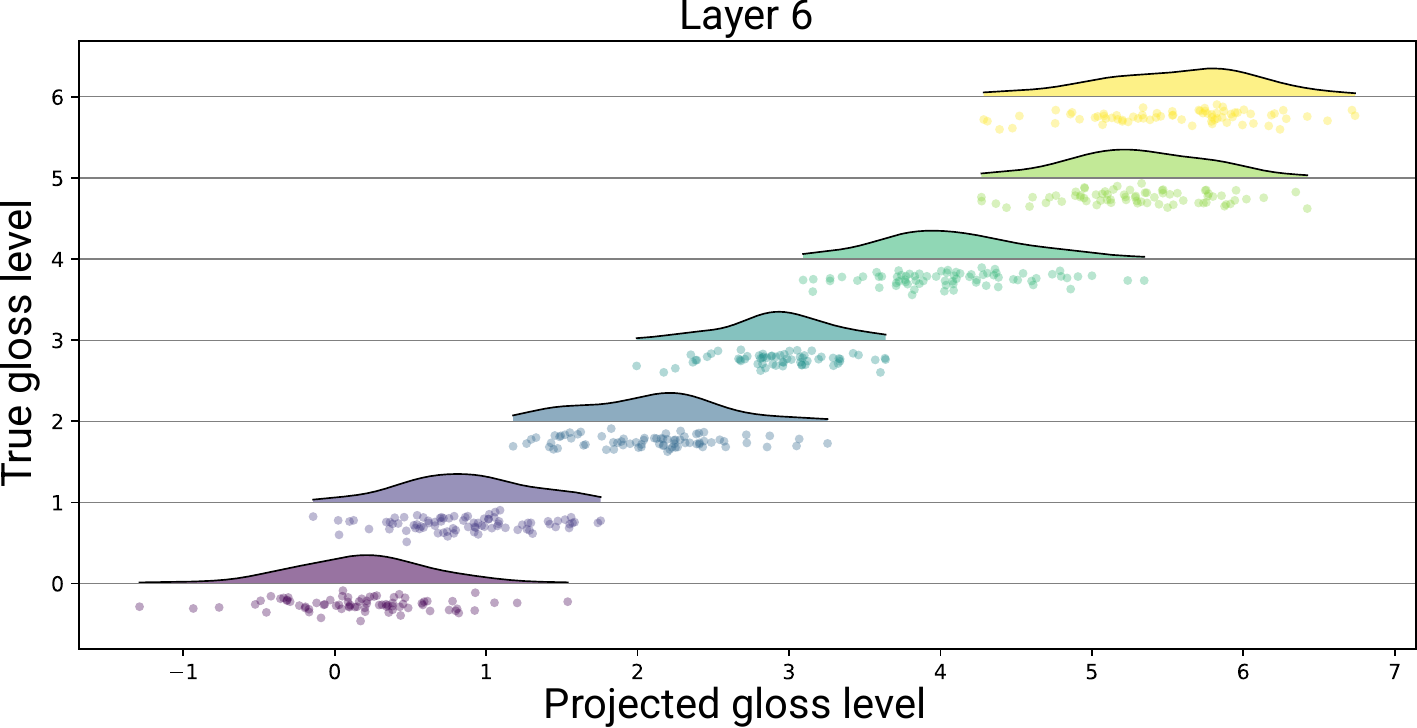}
    \caption{
        Projection of samples onto a one-dimensional space (horizontal axis) learned with a linear ridge regression model.  
        For all samples at each true gloss level (vertical axis), we show their scattering in the projected space, with small vertical jitter for visibility, together with the resulting half violin plot.
        Each point is projected using a model that was trained without that sample (cross-validation).
    }
    \label{fig:gloss}
\end{figure}
\begin{figure*}[t]
    \centering
    \includegraphics[width=\textwidth]{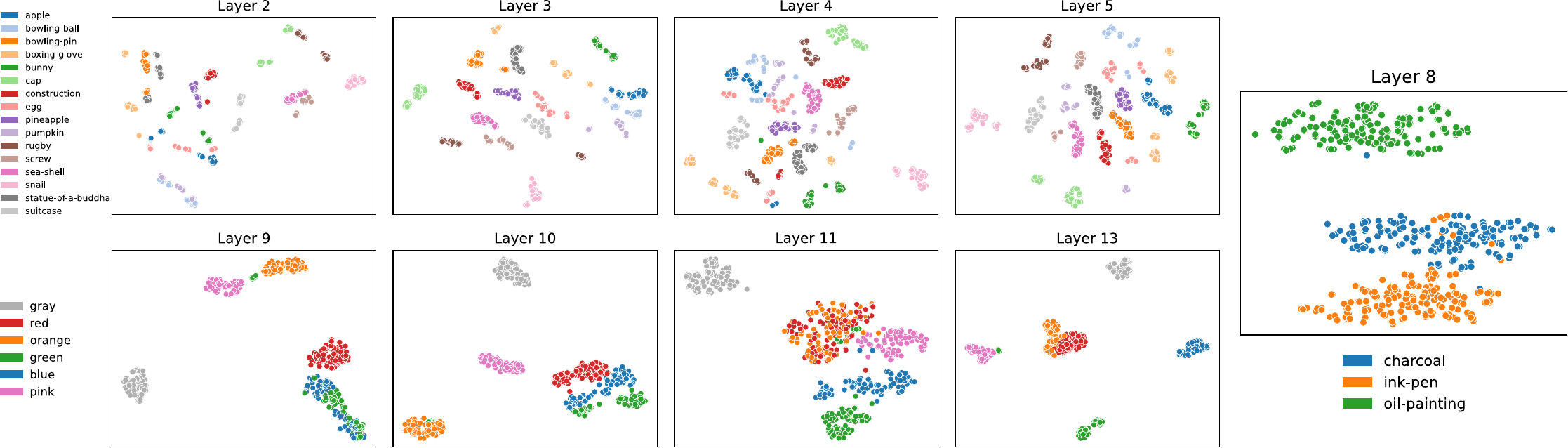}
    \caption{
       Representative 2D t-SNE visualizations illustrating the internal separation of samples within each label.  
\textit{Top:} Early layers colored by \emph{geometry} labels. \textit{Bottom:} Late layers colored by \emph{color} labels. \textit{Right:} Layer 8 colored by \emph{style} layers.
    }
    \label{fig:tsnes}
\end{figure*}

\subsection{Embedding of Categorical Factors}
\label{sec:embed_categorical}

After confirming that generative factors are hierarchically distributed across layers,
we next examine the internal structure of individual layers to assess intra-class disentanglement and information organization. For categorical factors (i.e., geometry, illumination, color, and style), we perform layer-wise t-SNE~\cite{vandermaaten08tSNE} projections of the 512-dimensional embeddings into a two-dimensional space for visualization, with points colored according to their corresponding labels. This unsupervised approach allows to evaluate not only whether a given layer encodes a particular factor, but also whether its internal representations are organized according to class labels. As illustrated in \autoref{fig:tsnes}, each layer exhibits internal clusters corresponding to different class instances. Notably, Layer 8 (\autoref{fig:tsnes}, right) achieves a clear separation of styles into three well-defined clusters corresponding with the three input styles. Complete t-SNE visualizations per layer and label can be found in the supplementary material (Sec.~S4.1).

\begin{figure}[t]
    \centering
    \includegraphics[width=\columnwidth]{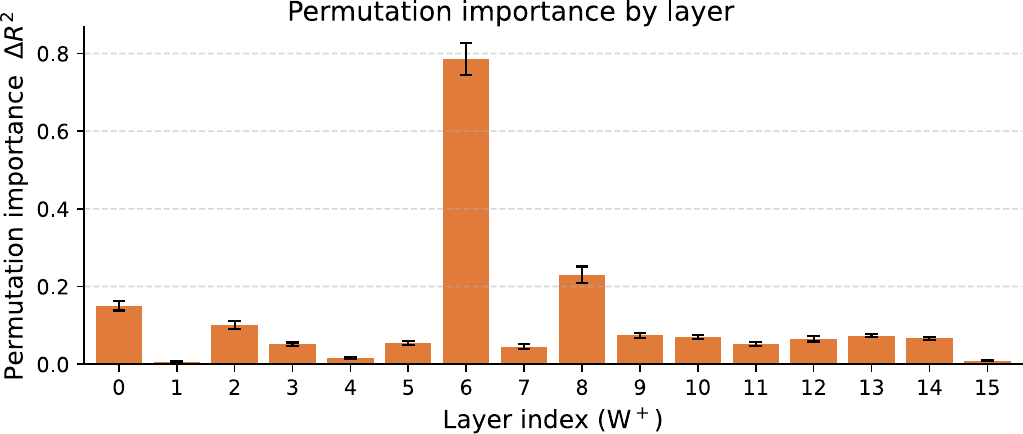}
    \caption{
        Uniqueness of gloss information per layer, measured by $\Delta R^2$ variations when predicting gloss labels without layer $i$. Please refer to the text for more details. 
    }
    \label{fig:infoloss}
\end{figure}

\subsection{Embedding of Gloss}
\label{sec:embed_gloss}

In this section, we examine how Layer 6 embeddings relate to the gloss labels in our dataset.
Unlike the categorical factors studied in \autoref{sec:embed_categorical}, gloss represents an ordinal variable, which makes the unsupervised t-SNE method unsuitable: t-SNE is designed to preserve local neighborhood relationships, but does not maintain the global ordering needed to assess gradual changes in gloss. We therefore employ alternative analyses to test whether the internal representation in Layer 6 reflects the expected progression of gloss levels.

As a first step, we apply supervised Ridge regression to predict gloss from the Layer 6 embeddings.
Using cross-validation, we train five separate models on different subsets of the training data and, for each model, project the held-out validation samples onto the resulting one-dimensional regression axis (samples are scattered uniformly along the $y$-axis in the plot for clarity). Each validation sample is always excluded from the training set of its corresponding model.
The results, shown in \autoref{fig:gloss}, reveal a well-defined linear trend with respect to gloss. Quantitatively, the resulting projections achieve a prediction MSE of $0.1224 \pm 0.0229$ against the ground truth and a Spearman correlation of $0.97$, confirming a strong monotonic relationship between the learned representation and the ground-truth gloss levels.

We next assess the unique contribution of each layer to the prediction of gloss. To this end, we first train a supervised linear model on all layers which learns to predict gloss labels. At test time, we evaluate the performance drop (by means of $R^2$ variation) when the embeddings of a given layer $i$ are made uninformative by randomly permuting them. A substantial decrease in performance when $w_i$ is permuted indicates that Layer $i$ contains unique information about gloss that cannot be compensated for by other layers,
meaning that this layer is explanatory of gloss. Consistent with previous visualizations, \autoref{fig:infoloss} shows that Layer 6 encodes the highest amount of unique gloss-related information.

Altogether, these results reveal a clear and interpretable encoding of gloss within the latent space, showing that this perceptual attribute can emerge spontaneously in an unsupervised model even when trained on painterly depictions with large variation in artistic styles.

\subsection{Ablation on the Number of Styles}
\label{sec:ablationstyles}

\begin{table}[t]
    \centering
    \caption{
    Quantitative results for the ablation study on the number of styles. Reported metrics represent compactness (\emph{Reg}, lower is better), and smoothness (\emph{PPL}, lower is better) of the latent spaces.
    }
    \label{tab:ablationmetrics}
        \begin{tabular}{lcccc}
        \toprule
        \#Styles & 2 & 3 & 4 & 5 \\
        \midrule
        Reg ↓ & 11.13 & 18.40 & 21.68 & 22.13 \\
        PPL ↓ & 54.32 & 54.03 & 60.22 & 55.20 \\
        \bottomrule
        \end{tabular}
\end{table}

In this section, we analyze the effect of varying the number of styles included in the dataset. We quantitatively assess the \emph{latent space compactness} as a proxy for disentanglement, utilizing the regularization metric (\emph{Reg}) proposed by Richardson et al.~\cite{richardson2021encodingpsp}.

We also compute the Perceptual Path Length (\emph{PPL}) which, since its introduction in the StyleGAN paper~\cite{karras2019stylesgan}, has been widely used to evaluate the \emph{smoothness and continuity} of latent spaces.

\autoref{tab:ablationmetrics} shows the result of training four different models with increasing number of styles. We observe a trend where smoothness and continuity of the latent spaces (measured by \emph{PPL}) remains steady despite variations in the number of styles, showing that the continuity property of the GAN-based pipeline is not affected by complexity increments in the dataset. 
On the other hand, we see how an increasing complexity in the dataset (here represented as a larger number of styles) naturally leads to a gradual and controlled decrease of the latent space compactness (higher \emph{Reg} values), which needs to account for the increased amount of information. This shows how the setup used in the work can be expanded to additional styles while keeping the \emph{disentanglement} and \emph{compactness} properties.
For additional details on the metrics used in this section, please refer to the supplementary material (Sec. S5).

Taken together, these analyses reveal a rich and interpretable organization of gloss and style within the latent space, showing that complex appearance cues can emerge without supervision.
This understanding shows the viability of using the information encoded in our rich latent space to control the image-synthesis application presented in \autoref{sec:diffusion}.

\section{Application: Style- and Gloss-guided Diffusion Pipeline}
\label{sec:diffusion}

We propose a diffusion pipeline that enables intuitive and controllable stylization of a reference image by leveraging the style- and gloss-aware latent space identified in \autoref{sec:analysis}.
To avoid the costly and time-intensive process of training diffusion models from scratch, recent research has explored combining the strengths of GANs (e.g. disentanglement, continuity, and latent space interpretability) with those of diffusion models (e.g. robustness, generalizability, and image quality)~\cite{li2024stylegan, gandikota2024concept}. A common strategy involves the use of lightweight adapters~\cite{ye2023ipadapterIPA}, which provide a balance between efficient training and effective conditioning by reusing the strong priors of pretrained models while incorporating novel cues.

\begin{figure}[t]
    \centering
    \includegraphics[width=\columnwidth]{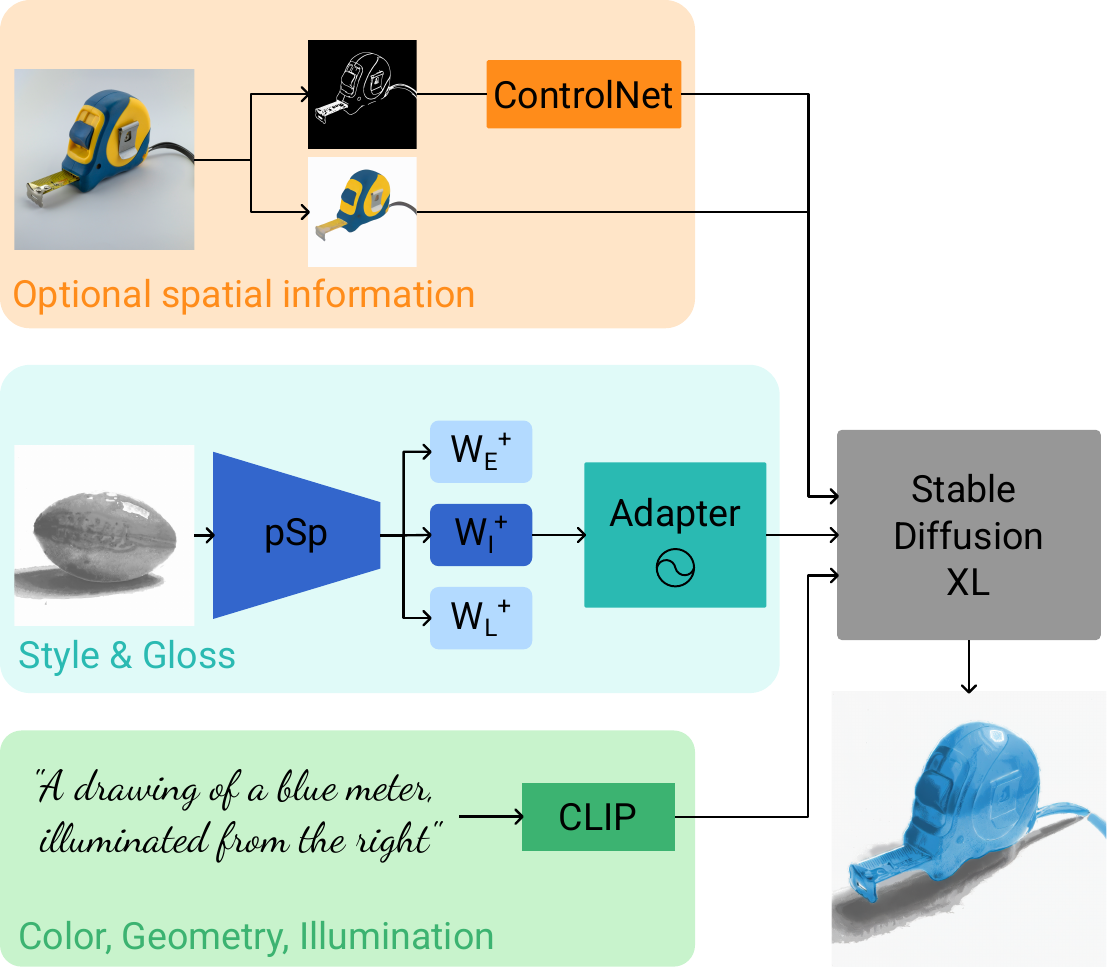}
    \caption{
        \textbf{Diagram of the proposed diffusion-based pipeline.} It can handle up to three inputs: (i) a text prompt defining the color, geometry, and illumination, (ii) a style and gloss reference image that uses the intermediate layers of the latent space created in previous sections, and (iii) an optional image to guide spatial information of geometry and albedo. 
    }
    \label{fig:diffpipeline}
\end{figure}

\begin{figure*}[t]
    \centering
    \includegraphics[width=\textwidth]{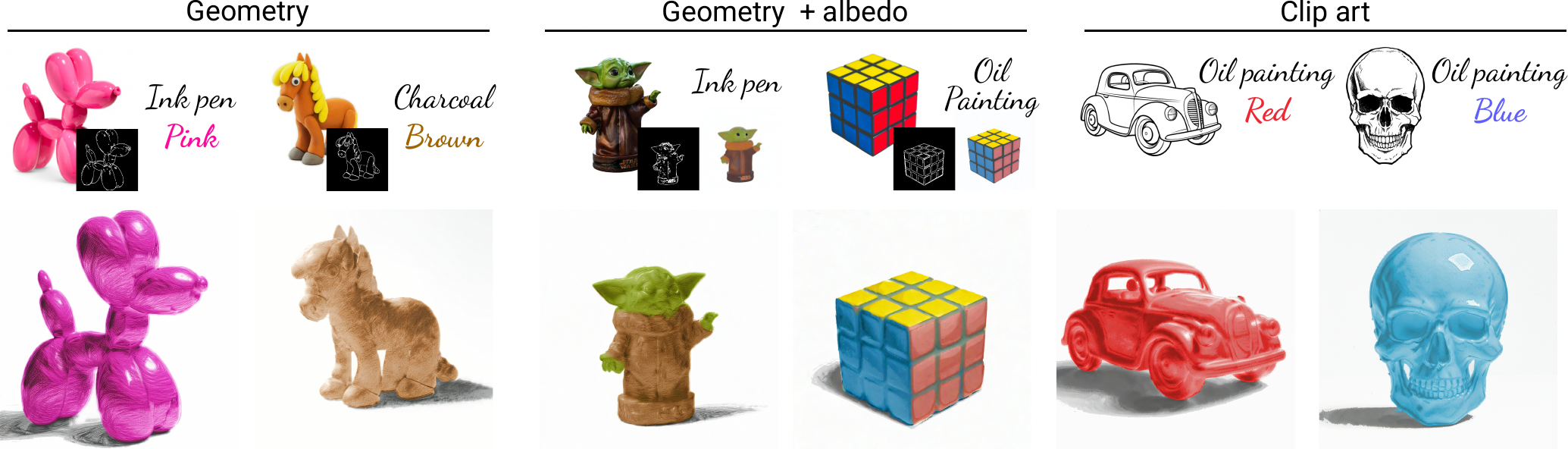}
    \caption{
        \textbf{Examples of three input modalities supported by our pipeline}. 
        Our method allows different levels of conditioning on geometry and color:  color can be specified through a text prompt (\textit{left}) or through spatial guidance using an albedo map (\textit{middle}).  
        Geometry is provided as edge or border maps, and by leveraging a ControlNet pretrained on Canny edges the pipeline also generalizes to other geometric representations such as clip-art drawings (\textit{right}).
    }
    \label{fig:spatial}
\end{figure*}

In our proposed pipeline, 
the adapter conditions generation on \emph{gloss} and \emph{style} features by leveraging the information encoded in the intermediate $W+$ embeddings ($W_I^+$); Our previous experiments have shown that these embeddings are
disentangled and highly informative of these two factors (\autoref{sec:latent_space}), making them well-suited for fine-grained control.
Conditioning the pipeline on the intermediate embeddings further validates the good properties of such representation, which needs to be properly disentangled and informative to result in a predictable and usable diffusion pipeline.
The text prompt conditions complementary factors such as \emph{geometry}, \emph{illumination}, and \emph{color}. This design allows the model to exploit the pretrained text encoder's prior knowledge, thereby generalizing to novel geometries or colors not observed during training of the adapter. To further enhance geometric control, we propose to integrate a pretrained ControlNet conditioned on Canny edges~\cite{zhang2023adding}. Finally, the user may optionally include an albedo map to specify the base albedo of the shape, providing finer-grained control over color. 
We base the implementation of our custom adapter on that of \textit{W+ Adapter}~\cite{li2024stylegan}, using Stable Diffusion XL 1.0 as backbone of the diffusion pipeline, and use as loss function the accuracy of the model when predicting the injected noise, as originally proposed by Rombach et al.~\cite{rombach2022high}.
We compare the performance of our model against state-of-the-art approaches, including (i) general-purpose text-to-image (T2I) models, (ii) style transfer methods, and (iii) Artist-Inator~\cite{subias2025artistinator}, a recent diffusion-based non-photorealistic stylization framework that is the closest related work to ours.
An overview of the designed pipeline is shown in \autoref{fig:diffpipeline}. Additional implementation details can be found in the supplementary material (Sec.~S2.2).

\subsection{Evaluation of Results}
\label{sec:evaluation_results}
This section presents qualitative and quantitative results of the proposed pipeline, including: (i) an evaluation of its different conditioning strategies, (ii) a comparison with prior work, and (iii) an analysis of the granularity when defining perceptual attributes such as gloss.

\noindent{\textbf{Conditioning via Text Prompt. }}
The minimal conditioning setup of the pipeline requires an input image for style and gloss guidance, and a text prompt for color, geometry, and illumination guidance. 
Current implementation allows defining one of the four illuminations used in the training dataset (\autoref{sec:dataset}) via text prompt (e.g., \textit{illuminated from the right}). Definition of color and geometry through text prompt can be made via free text, and results ultimately depend on the expressivity and accuracy of text embeddings, obtained with the pretrained text encoder used (in our case, CLIP~\cite{radford2021learning}).
Although versatile, defining geometry solely through text introduces variability proportional to the ambiguity of the described instance (e.g. the shape of \textit{Yoda} is more specific than that of \textit{toy}).
A Figure illustrating several examples generated under this text-based conditioning can be visualized in the supplementary material (Sec.~S6).
Results demonstrate that targeted modifications can be achieved intuitively by altering only specific parts of the text or image prompt, while leaving the other factors unchanged. 

\begin{figure*}[t]
    \centering
    \includegraphics[width=\textwidth]{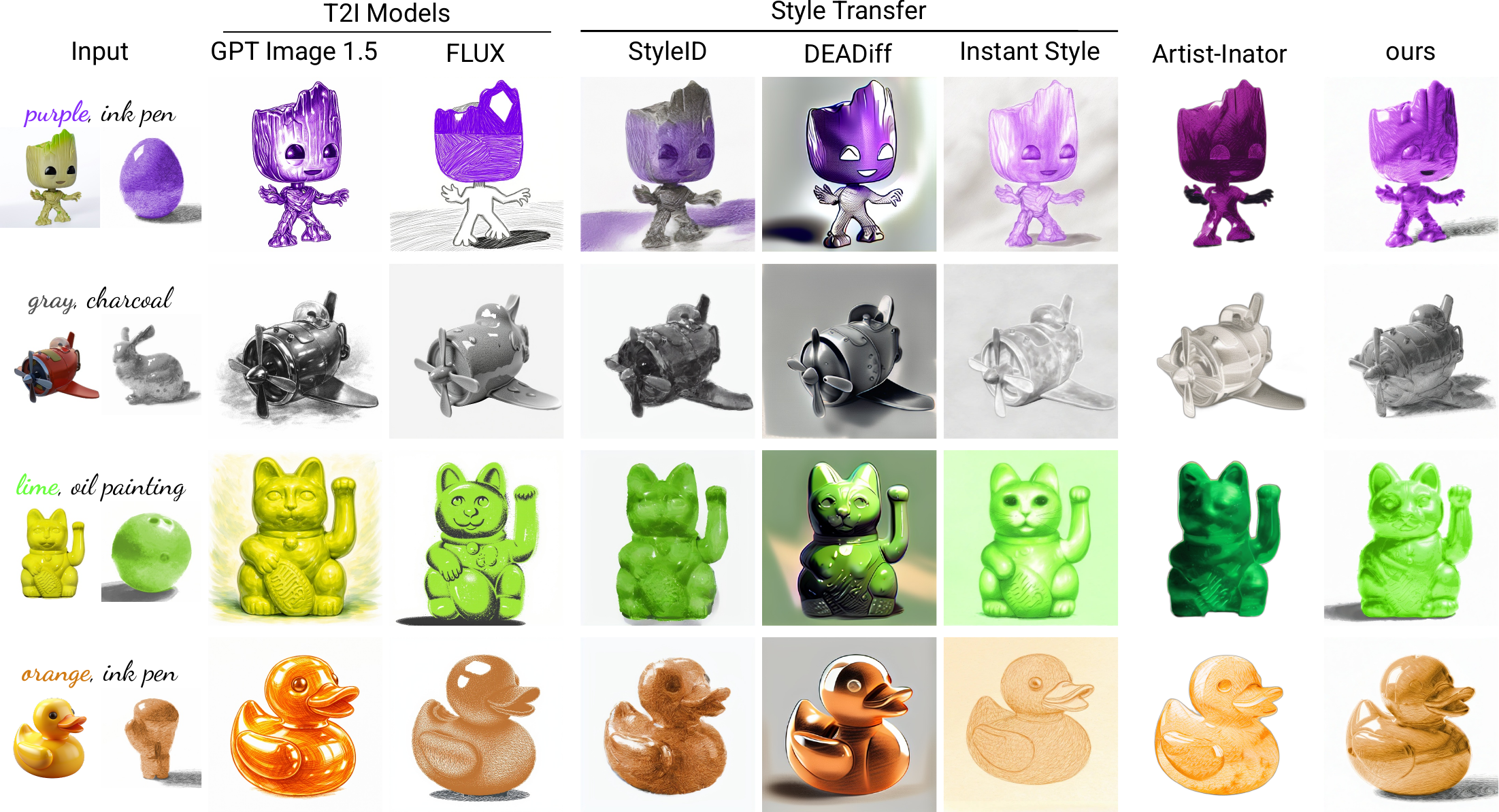}
    \caption{
        \textbf{Qualitative comparison with state-of-the-art models}.
        General-purpose T2I models produce visually striking images but often deviate from the target style and geometry. GPT Image 1.5 can often approximate the reference style,
        but its outputs offer only limited fine-grained control and may vary across prompts, making precise manipulation more challenging than with our method (\autoref{fig:traversalsSOTA}).
        Style-transfer methods better respect the reference style but tend to generate less compelling or less stable outputs.
        Our method achieves visually appealing results while maintaining high fidelity to both the reference geometry and the intended painterly style.
    }
    \label{fig:sota}
\end{figure*}

\begin{figure*}[t]
    \centering
    \includegraphics[width=\textwidth]{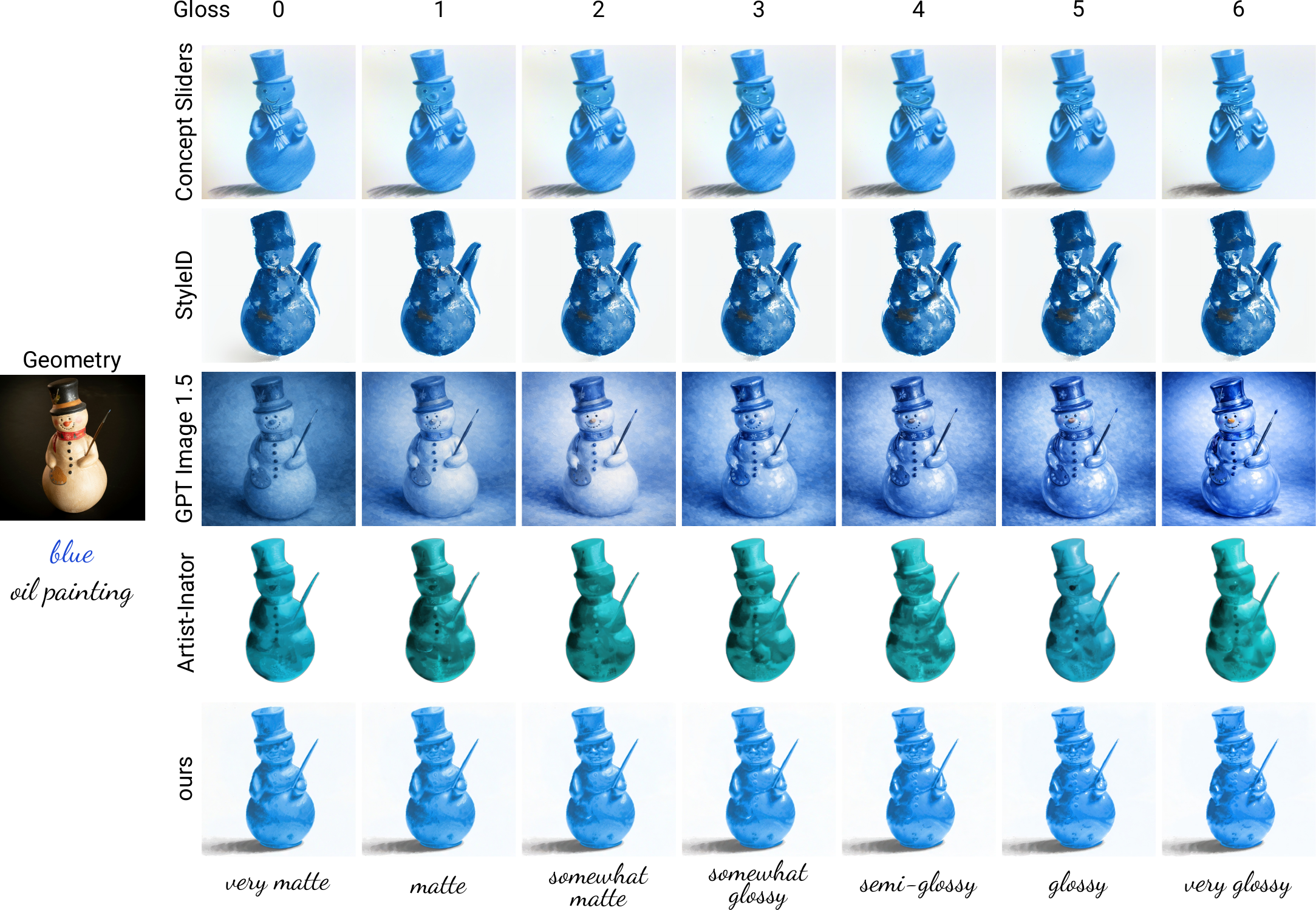}
    \caption{
        \textbf{Comparison of gloss control}. 
        We compare our method against a diverse set of baselines: ConceptSliders~\cite{gandikota2024concept}, a LoRA-based method; StyleID~\cite{Chung_2024_CVPR}, a training-free diffusion-based style-transfer method; GPT Image 1.5~\cite{openai_gpt4o_image2025}, a general-purpose image-generation model; and Artist-Inator~\cite{subias2025artistinator}, the closest prior work to ours, which is based on ControlNet.
        For our method, the target gloss level is controlled with a slider, shown at the top of the figure. For Artist-Inator and GPT Image 1.5, it is specified through a text prompt, shown at the bottom and ablated in Sec. S8 of the supplementary material. For StyleID and ConceptSliders, gloss is varied by traversing a hyperparameter (see Sec. S7.2 for details).
        Our approach produces a clear, gradual modulation of gloss while preserving geometry and overall style. In contrast, the baselines have limited ability to produce smooth, continuous gloss changes while maintaining identity and fine details. GPT Image 1.5, in particular, often produces visually striking results but struggles to maintain a consistent gloss progression.
    }
    \label{fig:traversalsSOTA}
\end{figure*}

\noindent{\textbf{Conditioning with Additional Spatial Information. }}
As illustrated in the pipeline architecture (\autoref{fig:diffpipeline}), additional spatial information can be introduced to control both shape and color. Shape is specified using a Canny edge map, which is then fed to a ControlNet~\cite{zhang2023adding}, while color is defined through an albedo map computed automatically with Marigold~\cite{ke2024repurposingmarigold}. The albedo map is applied during the late stage of the diffusion process to adjust the latent representations at inference time (additional details in Sec. S2.2). As shown in \autoref{fig:spatial}, this strategy enables precise control over the output geometry, effectively supporting style transfer with the learned styles. Additionally, the current configuration enables automatic generalization to diverse geometry representations, including clip-arts.

\noindent{\textbf{Comparison with Prior Work. }}
Image synthesis methods have attracted increasing attention due to the unprecedented quality achieved by recent generative models. We evaluate our approach against representative methods from four categories: %
\begin{itemize}
    \item General-purpose text-to-image (T2I) methods, which generate images directly from text prompts (with no \textit{style} reference), and are designed for a broad range of tasks. This category includes the methods of FLUX~\cite{flux2024} and GPT Image 1.5~\cite{openai_gpt4o_image2025} %
    \item Style transfer architectures, which specialize in transferring a reference style onto a target content image. We evaluate against StyleID~\cite{Chung_2024_CVPR}, DEADiff~\cite{qi2024deadiff}, and InstantStyle~\cite{wang2024instantstyle}.
    \item ConceptSliders~\cite{gandikota2024concept}, a method based on Low-Rank Adaptation (LoRA), which serves as a baseline for evaluating continuous gradients on the gloss dimension.
    \item Artist-Inator~\cite{subias2025artistinator}, a gloss-aware stylization method, which represents the closest approach to our work.
\end{itemize}

All models are guided on geometrical information using the same input image (\autoref{fig:sota}, first column), and style is defined either via an image input (\autoref{fig:sota}, second column), or via a textual description, depending on the method's requirements. The remaining factors of \textit{style}, \textit{gloss}, \textit{color}, and \textit{illumination} are defined via text prompt for T2I methods and Artist-Inator, while style transfer models use an additional reference image. %
Results in \autoref{fig:sota} reveal a trend in which general-purpose methods produce the most visually appealing outputs, at the cost of a lower adherence to the specific style. In contrast, style transfer approaches show reduced performance, which we attribute to domain shift: some of them rely on general-purpose image encoders that perform well with in-the-wild images, but are less effective with our non-photorealistic depictions of isolated objects on a white background. Finally, our method achieves strong adherence to both the reference style and gloss cues, while maintaining an adequate image quality.
Further details regarding the execution of each of the aforementioned methods and the design of the user study are provided in the supplementary material (Sec.~S7).

\subsection{Continuous Representation of Gloss}
A central strength of our approach lies in the precise control it offers over gloss.
Leveraging the good properties of our $W+$ space regarding continuity of the gloss attribute (\autoref{sec:embed_gloss}), we evaluate to what extent this property is preserved by the trained adapter and compare its performance with prior approaches. 
As shown in \autoref{fig:traversalsSOTA}, our model achieves the most faithful and continuous traversal of the gloss dimension, progressively adjusting highlights and specularities to increase perceived gloss while leaving other appearance factors unchanged.
Among competing approaches, 
ConceptSliders achieves an adequate control of gloss, but fails at keeping the identity of the reference geometry.
StyleID falls short in terms of visual quality, probably due to the domain shift.
GPT Image 1.5~\cite{openai_gpt4o_image2025} again delivers visually attractive outputs, but this comes at the cost of limited controllability over gloss.
Artist-Inator~\cite{subias2025artistinator} generally follows the gloss instruction, particularly at the extremes, but fails to produce a smooth progression.

\subsection{User Study}

We conducted a user study, composed of two experiments, to further validate the performance of our model. We compare the behavior of our model against relevant baselines using several setups that cover a wide variety of appearances. The stimuli were selected to span a representative range of object geometries, appearance attributes (color and gloss), and artistic styles, allowing us to evaluate the proposed method under diverse conditions while keeping the study manageable.
A total of 21 participants (mean age = 27.9 years; 7 female) took part in the study.

\vspace{1em}

\noindent\textbf{Experiment 1: Style transfer.} Participants were shown a style reference image, a content reference image, and seven candidate results generated by each of the evaluated methods shown in \autoref{fig:sota}. They were asked to rank the candidate images, via drag-and-drop, from worst to best according to the accuracy of the \textit{style transfer} from the style reference onto the content reference. Participants were allowed to freely refine their ordering until satisfied. The experiment was conducted across 12 examples spanning different objects, colors, artistic styles, and gloss levels. We report the following metrics:

\begin{itemize}
    \item Mean rank, computed as the average position assigned to each method across all examples and participants.
    \item Rank Product (RP), which reflects the overall user preference across all questions, and is computed as $RP(g) = (\prod_{k=1}^{K}r_{g,k})^{\frac{1}{K}}$, where $r_{g,k}$ represents the average ranking achieved by model $g$ in question $k$, and $K=12$. 
    \item Preference (Pref.) shows with what percentage our model was preferred over a certain alternative.
\end{itemize}

\begin{table}[t]
    \centering
    \caption{
    Quantitative results for Experiment 1 (Style transfer), comparing our method with alternative style transfer approaches. We report the mean rank with standard deviation, the Rank Product (RP), and the preference rate of our method over each alternative (Pref.). Lower mean rank and RP indicate better performance, while higher preference indicates better performance of our model.
    }
    \label{tab:userstudy1}
    \begin{tabular}{lccc}
        \toprule
        \textbf{Method} & \textbf{Mean Rank ↓} & \textbf{RP ↓} & \textbf{Pref. ↑} \\
        \midrule
        \textit{Ours} & $\mathbf{1.88 \pm 1.46}$ & \textbf{1.534} & -- \\
        InstantStyle & $3.23 \pm 1.39$ & 2.919 & 80.95\% \\
        StyleID & $4.02 \pm 2.25$ & 3.274 & 76.59\% \\
        GPT Image 1.5 & $4.23 \pm 1.86$ & 3.713 & 85.71\% \\
        Artist-Inator & $4.33 \pm 1.51$ & 4.036 & 87.70\% \\
        FLUX & $4.69 \pm 1.70$ & 4.273 & 87.30\% \\
        DEADiff & $5.61 \pm 1.43$ & 5.370 & 93.65\% \\
        \bottomrule
    \end{tabular}
\end{table}

The results of Experiment 1 (\autoref{tab:userstudy1}) show that participants consistently preferred our method over the alternatives in terms of \textit{style transfer accuracy}. Note that this experiment did not evaluate overall visual appeal, under which GPT Image may have achieved higher rankings.

\begin{figure}[t]
    \centering
    \includegraphics[width=\columnwidth]{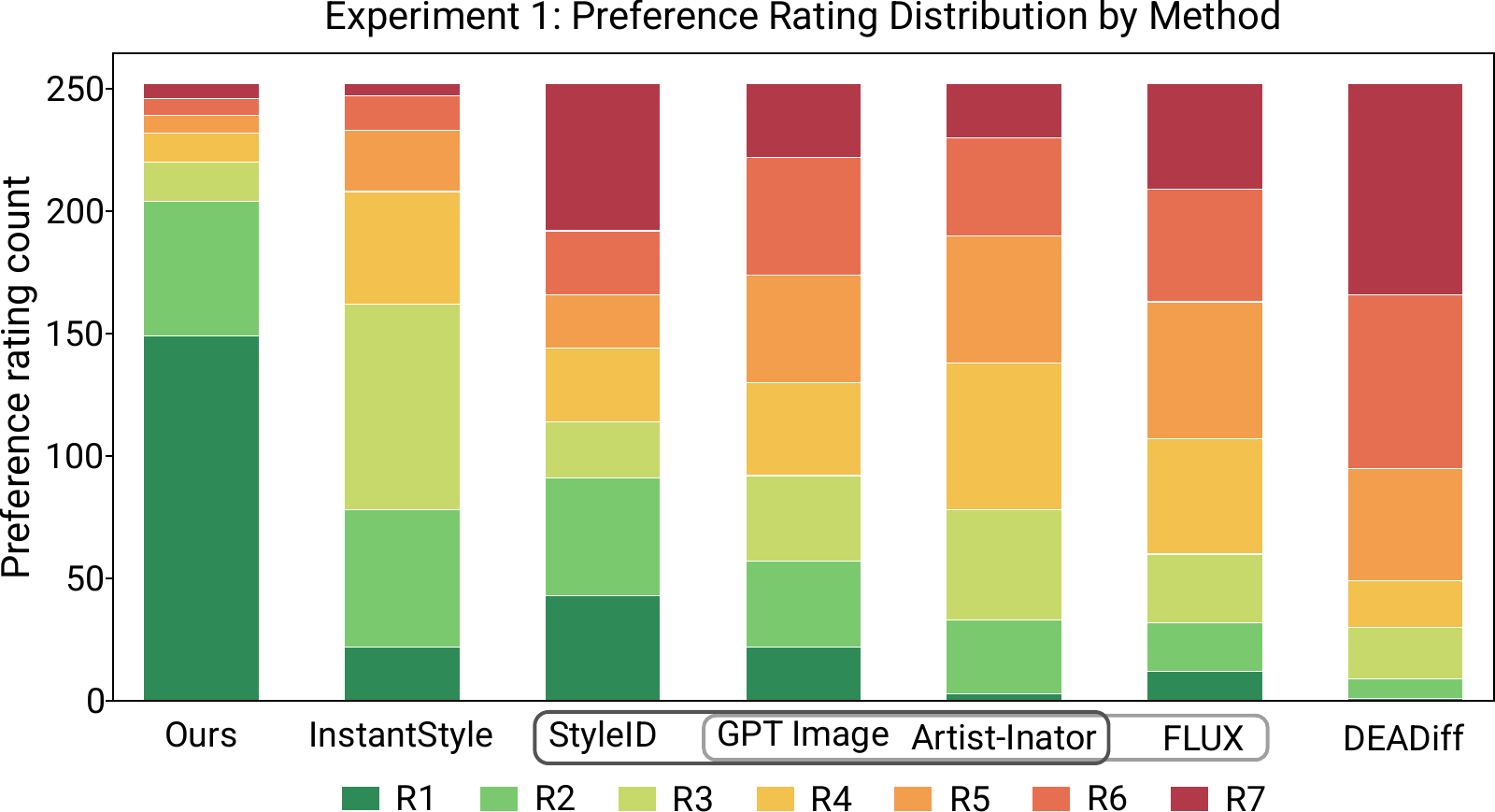}
    \caption{
        Distribution of preference rankings assigned to each evaluated method in Experiment 1, aggregated across all participants and the 12 evaluation examples. Each stacked bar shows the total number of times a method received each ranking position (R1–R7). Pairwise statistical comparisons reveal significant differences between methods, except for those enclosed in gray boxes, which are not statistically distinguishable.
    }
    \label{fig:exp1ranks}
\end{figure}

\autoref{fig:exp1ranks} further illustrates the distribution of rankings assigned to each method in Experiment 1, aggregated across participants and examples.
We further evaluated the ranking distributions performing Kruskal–Wallis tests followed by Bonferroni-corrected post-hoc comparisons \cite{wang2022tone,jarabo2014people} to determine whether differences in rankings were statistically significant. Numerical results are provided in Sec. S7.5 of the supplementary material.
Our method receives first-place rankings substantially more frequently than the alternatives, whereas lower-performing methods accumulate more low-ranking assignments and exhibit greater variability across ranking positions.
Consistent with \autoref{tab:userstudy1}, the results indicate a clear preference for our approach in terms of style transfer accuracy.

\vspace{1em}

\noindent\textbf{Experiment 2: Gloss odering}. We evaluate how accurately four methods (ConceptSliders, GPT Image, Artist-Inator, and ours) represent continuous variations along the gloss dimension. Participants were asked to arrange, via drag-and-drop, five images generated by each method from matte to glossy across five evaluation examples, and were allowed to freely refine their ordering until satisfied. We assess the accuracy of these orderings using the following metrics:

\begin{itemize}
    \item Absolute error between the user-ranked and ground truth gloss levels.
    \item RMSE between the user-ranked and ground truth gloss levels.
    \item Exact-match accuracy, defined as the percentage of times an image was ranked exactly on its ground truth level.
\end{itemize}

The results of Experiment 2 (\autoref{tab:ordering}) show that the gloss traversals generated by our model were ranked more accurately than those of the competing methods, achieving both the lowest mean error and the highest exact-match accuracy.

For conciseness, we include only the most representative analyses here. Additional details on the experimental configuration and execution, further analyses, disaggregated visualizations, and the statistical analysis of Experiment 1 are provided in Sec. S7 of the supplementary material.

\begin{table}[t]
    \centering
    \caption{
    Quantitative results for Experiment 2 (Gloss ordering) across all trials. We report the mean absolute error with standard deviation, RMSE, and exact-match rate. Lower error and RMSE indicate better performance, while higher exact rate indicates better performance.
    }
    \label{tab:ordering}
    \begin{tabular}{lccc}
        \toprule
        \textbf{Method} & \textbf{MAE ↓} & \textbf{RMSE ↓} & \textbf{\%Exact ↑} \\
        \midrule
        \textit{Ours} & $\mathbf{0.26 \pm 0.38}$ & \textbf{0.37} & \textbf{0.79} \\
        GPT Image 1.5 & $0.42 \pm 0.46$ & 0.58 & 0.66 \\
        ConceptSliders & $0.90 \pm 0.63$ & 1.18 & 0.48 \\
        Artist-Inator & $0.92 \pm 0.59$ & 1.21 & 0.43 \\
        \bottomrule
    \end{tabular}
\end{table}

\begin{figure*}[t]
    \centering
    \includegraphics[width=\textwidth]{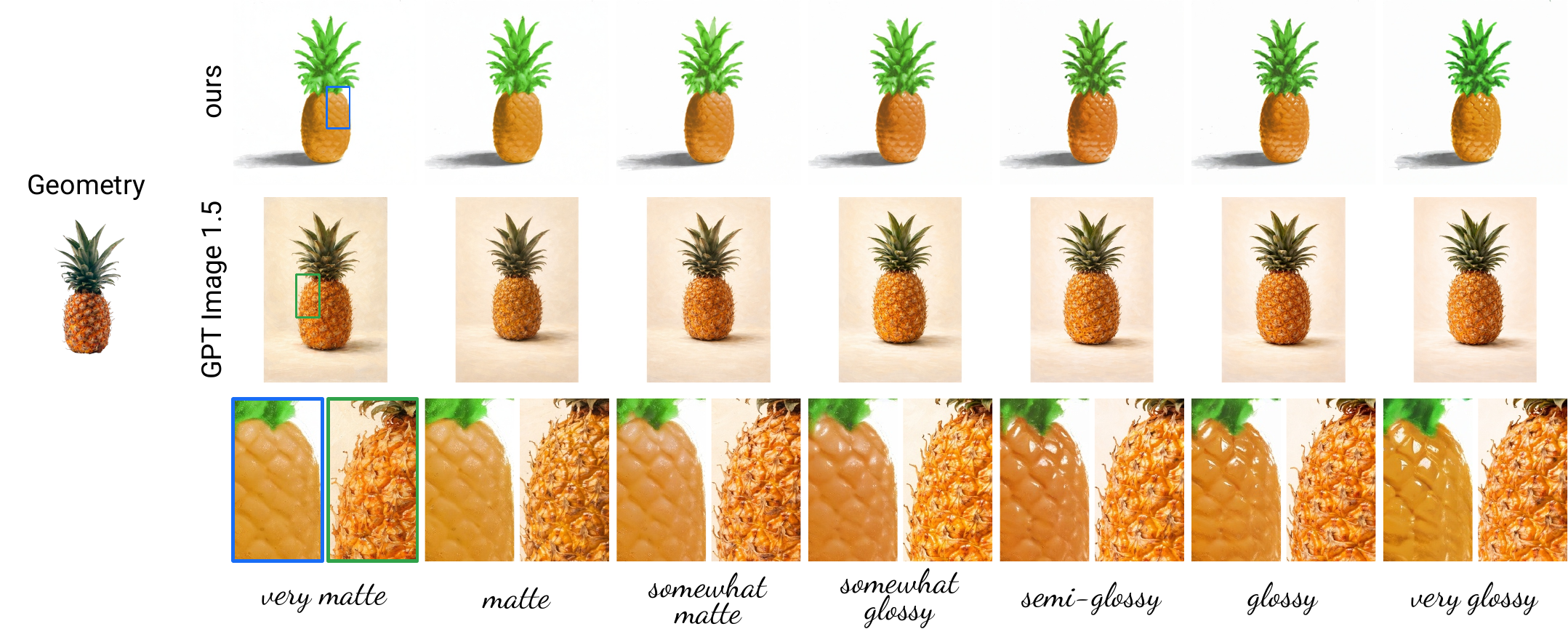}
    \caption{
        \textbf{Bias in general-purpose models.} Example illustrating how learned priors in large-scale text-to-image models can conflict with user prompts. In this case, GPT Image 1.5 struggles to generate glossy pineapples and instead introduces pineapple-like surface textures as gloss increases. In contrast, our disentangled representation enables more precise and continuous control over gloss while preserving the underlying object appearance.
    }
    \label{fig:gptimg_priors}
\end{figure*}

\section{Conclusions}

We presented a comprehensive analysis of how perceptual factors naturally organize within a hierarchical GAN-based architecture. Focusing on non-photorealistic imagery, we examined the interdependencies between artistic styles and other perceptual attributes, with a particular emphasis on gloss. Our analysis revealed a latent space that disentangles style from gloss, 
showing that a generative model can separate such factors without relying on explicit labels.
Building on this representation, we trained a lightweight adapter that enables the generation of non-photorealistic images of arbitrary objects with fine-grained control over style and gloss, while supporting flexible conditioning of geometry, color, and illumination through text and image references. %

\subsection{Limitations and Future Work}

Our proposed approach does have certain limitations.
First, we restricted the number of styles to three --charcoal, ink pen, and oil painting--, choosing them to be representative of a variety of hand-drawn or hand-painted styles. 
As a result, our model cannot natively generalize to unseen styles such as soft crayon or watercolor; handling these would require explicit training or fine-tuning. 
Nonetheless, the modular training and analysis framework used here could be easily extended to incorporate additional styles in future work. 
A second limitation relates to the highlight stylization. In photorealistic rendering, shiny reflections follow a closed-form appearance model and can be estimated directly. In artistic depictions, however, artists may use diverse conventions to convey glossiness, such as cross-hatching in ink drawings or white ellipses in cartoons. Our work focuses on a specific depiction of glossiness inspired by its photorealistic counterpart. Supporting more free-form artistic styles would require a different dataset-generation procedure, and the learning method would likely need some supervision to achieve comparable disentanglement.

Regarding the degree of control over the color, in the current implementation the albedo map is applied during late stages of the diffusion process (Sec. S2.2); while it works well generally, this strategy can lead to the loss of color information, particularly in fine details.
ControlNets~\cite{zhang2023adding} are a well-established approach to condition image generation on spatial information (such as albedo maps). However, to the best of our knowledge, no validated pretrained ControlNet for albedo maps is currently available for Stable Diffusion XL, and this therefore remains future work. 
Another interesting line for future work would be to combine traversals in our disentangled $W+$ space with those in a more general-purpose space such as CLIP~\cite{guerrero2024texsliders, cheng2025marble}, potentially combining their conditioning capabilities.
It could be interesting to study how to leverage the current expressive general-purpose models to create a varied non-photorealistic dataset that would be harder to generate with more traditional approaches (\autoref{sec:dataset}).

Finally, although the user study presented in this work is sufficient to support overall comparisons between methods, it was not specifically designed to analyze performance separately for each factor (e.g., artistic style or color). A larger-scale study would be required to draw statistically robust conclusions at this finer level of analysis.

\subsection{Discussion}

While prior analyses of latent spaces focus on understanding emergent perceptual factors~\cite{liao2023unsupervised}, they do not demonstrate whether they can be used to effectively control generation. Conversely, prior NPR-oriented generative work (notably Artist-Inator~\cite{subias2025artistinator}) provides stylization capabilities but does not yield a disentangled representation of perceptual factors and thus cannot offer predictable factor-wise control. Our work bridges these two directions: we introduced a style- and gloss-aware latent space, and show that exploiting this space leads to improved control and better style preservation, as confirmed by our user study.
Finally, as discussed in \autoref{sec:evaluation_results}, while our method may lack the generative capacity of large-scale general-purpose models such as GPT Image 1.5, it provides a suitable alternative in scenarios where precise control over generative factors is required. Furthermore, these large-scale models are typically proprietary and trained on undisclosed datasets, making it difficult to understand the causes of their successes and failures in specific scenarios.

The increased controllability of our method becomes particularly evident in individual generations. General-purpose models such as GPT Image 1.5 can often distinguish broad categories such as \textit{matte} and \textit{glossy}, yet struggle to reproduce fine-grained variations within these appearance attributes (\autoref{fig:traversalsSOTA}). This limitation may stem from the complex interactions between multimodal tokens in large generative models. Similar issues can also emerge when user prompts conflict with learned priors, as illustrated by the \textit{glossy pineapple} example in \autoref{fig:gptimg_priors}. In such cases, methods with stronger disentanglement between content and style, such as ours, provide more reliable control over the generated appearance.
Moreover, the disentangled structure of the latent space enables a higher degree of interpretability by directly inspecting the latent representation driving the generated image, an aspect that remains largely inaccessible in general-purpose models such as GPT Image 1.5.

Recent advances in generative image synthesis have reached unprecedented levels of visual fidelity, to the point that generated images can be indistinguishable from real photographs to the naked eye. However, these capabilities often rely on models with billions of parameters, which significantly hinders the explainability of their outputs. Interpretability constitutes an important research direction in the study of large-scale generative models, and 
we hope this work inspires new developments on controllable generative models and their relation to human visual perception, leading to improved content authoring tools. Our trained model, code and dataset are available at \url{https://graphics.unizar.es/projects/JimenezNavarro2026-NPRdisentanglement/}.

\section*{Acknowledgements}
This work has been supported by grant PID2022-141539NB-I00, funded by MICIU/AEI/10.13039/501100011033 and by ERDF, EU.
This work has also received funding from the Gobierno de Aragon through the project “HUMAN-VR” (PROY\_ T25\_24),
and from the European Union (ERC grant number 101220555, PROXIE). Views and opinions expressed are however those of the author(s) only and do not necessarily reflect those of the European Union or the European Research Council Executive Agency. Neither the European Union nor the granting authority can be held responsible for them. 
Additionally, Santiago Jimenez-Navarro was supported by a Gobierno de Aragon predoctoral grant (year 2025-2029).
Special thanks to the reviewers and to everyone who participated in the user study.

\bibliographystyle{elsarticle-num}
\bibliography{main}

\end{document}